\documentclass[12pt]{iopart}
\usepackage[dvips]{graphicx}

\begin{document}

\title[Single Member Selection in Ensemble Forecasting]{Single Member Selection in Ensemble Forecasting}

\author{F J Tapiador$^1$ and R Verdejo$^2$}

\address{$^1$Universidad de Castilla-La Mancha (UCLM),45071 Toledo, Spain}
\address{$^2$Swiss Federal Laboratories for Materials Testing and Research (EMPA), 9014 St Gallen, Switzerland}
\ead{francisco.tapiador@uclm.es}

\begin{abstract}

Ensemble forecasting is a technique devised to palliate sensitivity to initial conditions in nonlinear dynamical systems. The basic idea to avoid this sensitivity is to run the model many times under several slightly-different initial conditions, merging the resulting forecast in a combined product. We argue that this blending procedure is unphysical, and that a single trajectory should be chosen instead. We illustrate our case with a climate model. While most of the current climate simulations use the ensemble average technique as merging procedure, this paper shows that this choice presents several drawbacks, including a serious underestimation of future climate extremes. It is also shown that a sensible choice of a single estimate from the ensemble solves this problem, partly overcoming the inherent sensitivity to initial conditions of those non-linear systems with a large number of degrees of freedom. 
 
\end{abstract}
\maketitle

\section{Introduction}
A numerical model is said to present sensitivity to initial conditions (SIC) when no matter how close two initial conditions are they diverge as the system evolves \cite{Lorenz63}. This behavior arises from nonlinear processes: small initial differences are multiplied and grow to high magnitude orders, soon resulting in uncorrelated trajectories for arbitrarily-close initial points \cite{Ehrendorfer94}. SIC is one of the characteristics defining chaotic behavior \cite{Whitfield05}.

Climate models present this sensitivity \cite{Trevisan98} \cite{Tsonis01} \cite{Orrell05}. After a few iterations, a model may generate different forecasts for different albeit close initial conditions, making prediction very difficult \cite{Buizza05}. Ensemble forecasting methods have been proposed to circumvent this problem \cite{Hansen02} \cite{Kalnay03}. The key idea is to make the system evolve under several slightly-different initial conditions mimicking our uncertainty on the true initial conditions \cite{Vialard05}. By using many initial conditions we obtain a bunch of trajectories in phase space. Since the initial conditions are carefully chosen as to encompass the full dynamics of the system \cite{Anderson96}, it is sensible to assume that the true trajectory should lie within this set \cite{Sivillo97} \cite{Kalnay98}. In ensemble literature a trajectory in phase-space from a given initial condition is called a member of the ensemble. 

Climate ensembles provide two major outcomes. First, they give an estimate of the predictability of the forecast: if most of the members present a close evolution the system is considered highly predictable, and unpredictable otherwise \cite{Trevisan01} \cite{Toth01}. The second result is an estimate of the future climate, which is given by a combination of the forecast of the individual members \cite{Houtekamer97} \cite{Cheung01}. 

Ensemble methods are aimed to provide a probabilistic forecast -that is, a prediction described by probabilistic distribution functions (pdfs)- but the average of the members is widely used as an estimate of the mean future climate: \cite{Tracton93} \cite{Toth97} \cite{Shukla98} \cite{Pan98} \cite{Cai03} \cite{Ananos04} \cite{Kong06}. The reason is that the average is expected to have less error covariance than any single member: if $\textbf{u}_0$ is the anomaly of the true trajectory of zero mean, $\hat{\textbf{u}}$ is the anomaly of a single forecast, and $\bar{\textbf{u}}$ is the anomaly of the ensemble average of $m$ members, we have from \cite{Leith74} that:
\begin{eqnarray}
\begin{array}{l}
\left\langle  (0-\textbf{u}_0)(0-\textbf{u}_0)^{T} \right\rangle \stackrel{t \rightarrow \infty}{\rightarrow}  U \\
\left\langle  (\hat{\textbf{u}}-\textbf{u}_0)(\hat{\textbf{u}}-\textbf{u}_0)^{T} \right\rangle \stackrel{t \rightarrow \infty}{\rightarrow}   2 U \\
\left\langle  (\bar{\textbf{u}}-\textbf{u}_0)(\bar{\textbf{u}}-\textbf{u}_0)^{T} \right\rangle \stackrel{t \rightarrow \infty}{\rightarrow}   (1+ \frac{1}{m}) U
\end{array} 
\end{eqnarray}

Which proves that, in the infinity limit, the ensemble average has less error covariance than a randomly-chosen single forecast. 

In the first part of this article we will show that this averaging procedure results both in considering phase state points incompatible with the dynamics of the system, and in producing a merged forecast that does not necessarily capture the extreme values of the forecast. In terms of future climate, this means that ensemble averages might be underestimating climatological extremes. We will use a simple numerical example (a logistic map) and an actual climate ensemble to illustrate the effects of ensemble averaging. In the second part we will provide an alternative to ensemble averaging.

\section{Ulam's map example}

Let us use the logistic map with parameter $\mu$, $x_{n+1}= 1-\mu \cdot x_{n}^2, \mu =2 $ (Ulam's map) to describe the ensemble averaging procedure. This system is widely used in the study of chaos dynamics. The support of Ulam's map is the [-1,1] interval, being dense in this interval and having two stable points at $x=-1$ and $x=1/2$. The map is ergodic and mixed, and its natural invariant density \cite{Beck93} is $\rho (x) = \frac {1}{\pi (1-x^2)^{1/2}} $ (Fig. 1). The chaotic behavior of the map dynamics can be established by calculating the Lyapunov exponent, resulting in $log(2)$. Since this value is positive, the system shows SIC. As the map is ergodic, $\rho (x)$ can be estimated either as ensemble or time average using an histogram function.

To investigate the properties of the ensemble average, we can construct an ensemble around a non-stable initial point such as $x_0=0.3$. We will not assume that we know the error distribution around the true initial value. Instead, we will take uniformly-distributed ICs $x_i \in [x_0-10^{-8}, x_0+10^{-8}]$ (a more careful choice \cite{Bark98} would make no difference in this example). Figure 2 shows how the ensemble average of Ulam's map compares with a true trajectory after 100 iterates (RK4 scheme). The trajectory of the ensemble average bears little resemblance to the actual behavior of any possible trajectory such as the reference orbit shown, being clear from the figure that if the ensemble average yields a high correlation it is because it remains almost equidistant to the  $\left\{-1,1\right\}$ extremes. 

If we calculate the mean value of the average after 10,000 iterations, it compares well with the true value (-0.0004 vs. 0.0). The variance, however, is 0.00633, which  is well below the actual value (0.5). The same would apply for the derived tent map $x_{n+1}=2 \cdot x$  for $x \leq 1/2$ and $2 \cdot (1-x)$ for $x > 1/2$, with support [0,1], even when the natural invariant density of this map is constant $\rho(x)=1$. After 10,000 iterations, the variance of the averaged ensemble is 0.05, compared with 0.72 for an typical orbit. 

These examples show that the ensemble average may give a good statistical $r^2$ correlation in the long term, but we are building up a trajectory without a physical counterpart. Taking the ensemble average as a representative of the true dynamics of the system does not guarantee capturing the ensemble behavior: there is no initial value for Ulam's map capable of generating a trajectory similar to the one described by this average. In fact, Ulam's map contra example shows that by using the ensemble average we could end up selecting what actually is the less likely state of the system as the system's dynamics. Any orbit of this map spend most of the time near the extreme points $\left\{-1,1\right\}$, while the ensemble average wrongly suggests a zero mode.  

\section{Ensemble averaging in climate models}
To understand why the ensemble average gives less error covariance than a randomly-chosen single forecast $\hat{\textbf{u}}$ consider that: 
\begin{equation}
\left\langle ( \hat{\textbf{u}}-\textbf{u}_0)(\hat{\textbf{u}}-\textbf{u}_0)^{T} \right\rangle = \left\langle \hat{\textbf{u}} \hat{\textbf{u}}^T + \textbf{u}_0 \textbf{u}_0^T - \hat{\textbf{u}} \textbf{u}_0^T - \textbf{u}_0 \hat{\textbf{u}}^T     \right\rangle = 2 U
\end{equation}

Since the last two terms in the rhs vanish as $t \rightarrow \infty $ as the system becomes uncorrelated in time, $\left\langle \textbf{u}_0 \textbf{u}_0^T \right\rangle$ is the true error covariance $U$, and $\left\langle \hat{\textbf{u}} \hat{\textbf{u}}^T  \right\rangle= U$ as $t \rightarrow \infty $.

If, however, we take $ \hat{ \textbf{u}}= \bar{ \textbf{u}}= \frac{1}{m} \sum_{i}{ \textbf{u}_i} $, then 

\begin{equation}
\left\langle \hat{\textbf{u}} \hat{\textbf{u}}^T  \right\rangle = \frac{1}{m} \sum_{i}{ \textbf{u}_i} \frac{1}{m} \sum_{j}{\textbf{u}_j} \stackrel{t \rightarrow \infty}{\rightarrow}  \frac{1}{m} U \\   
\end{equation}

Thus,
\begin{equation}
\left\langle  (\hat{\textbf{u}}-\textbf{u}_0)(\hat{\textbf{u}}-\textbf{u}_0)^{T} \right\rangle \stackrel{t \rightarrow \infty}{\rightarrow}  (\frac{1}{m}+1) U 
\end{equation}

The result being that the performance of the ensemble average is due to the $1/m$ factor in the average. From Eq. 3 also follows that the larger the number of ensemble the lower the error covariance. 

Nonetheless, the same reasoning could be applied to any linear function in the form $ \tilde{ \textbf{u}}= \frac{1}{m^n} \sum_{i}{ \textbf{u}_i}, n \in \Re $. These \emph{means} would be also unbiased estimates of the true state $\textbf{u}_0$ and also unbiased estimates of the mean of the forecast, decreasing the corresponding error covariance of the $\left\langle \bar{\textbf{u}} \bar{\textbf{u}}^T \right\rangle$ term by a $\frac{1}{m^n}$ factor. These other estimates would provide even lower covariance than the average, but, as the average, none of them meet the basic physical requirement of being a possible state of the system. Ensemble average is thus a statistical way of reducing the covariance, but lacks a physical interpretation. 

An additional argument against ensemble averaging arises from probability theory. As mentioned, ensemble methods are aimed to provide probabilistic forecasts for future climate. It is known that global precipitation fields present a lognormal distribution \cite{Kedem94}. An ensemble average of precipitation forecasts would then be a weighted sum of lognormal pdfs. Application of the central limit theorem \cite{Feller71} proves that the sum of those lognormally-distributed pdfs converge in distribution to the normal (Gaussian) pdf. (The Lyapunov condition to apply the theorem holds for climate ensembles). The consequence is that the averaging procedure does not conserve the pdf of the members, but creates a new pdf with a different function. This effect is apparent in climate ensembles averages, as Figure 3 illustrates. Here, the average of the nine members of a global precipitation seasonal ensemble \cite{Palmer04} is compared with validation data (ERA-40 reanalysis), and with the objective best member (The objective best/worse member is defined as that member presenting the best/worse $r^2$ correlation against ERA-40). The ensemble technique is not intented to provide such daily fields estimates, but the figure is used here just to graphically illustrate what is happening with the pdfs when averaging. 

From figure 3 it is apparent that the ensemble mean shows a smoother field which is not representative of the original pdf of the members, and does not resemble the true precipitation field (ERA-40). In the ensemble average figure the high precipitation rates predicted by the ensemble members in the Pacific disappear, and areas of scarce precipitation such as California present artificially high estimates due to the averaging. 

The ensemble average provides good statistics in terms of $r^2$ correlation (Table 1, second column), but the maximum precipitation is severely underestimated (Table 1, third column). The spatial variance of the field is also lower than that for any member, as consequence of compensating dissimilar estimates for the same geographical area. More importantly, the ensemble mean does not necessarily reflect a feasible state of Earth's daily precipitation field. Figure 4 shows that the pdf of the ensemble average greatly differs from the members pdfs, with a severe underestimation of high precipitation values, and a reduction in light precipitation in favor of moderate precipitation rates. This normalization of the true pdf makes the averaging procedure unsuitable for predicting climate extremes. 

This is specially visible by comparing the characteristics of the objective best member (member 5) with the ensemble mean. Member 5 appears as a feasible global precipitation field whereas the ensemble average does not represent any likely state of the atmosphere. As in Ulam's map, the average of this ensemble has not other physical interpretation than being a derived quantity of otherwise physically-consistent estimates.  

One alternative to this problem is not averaging the many simulations but the statistics derived from the individual members. In this case, the ensemble mean is simply discarded, and the spread of the ensemble is used as an estimate of the limits of the future state. The drawback of this procedure, however, is that nothing can be said on the likely dynamics of the system. 


\section{Single member selection}
Even when ensemble averaging does not provide a physically-consistent trajectory, ensemble forecasting is a way of tackling SIC in dynamical systems. An alternative to the average is selecting a single member of the ensemble. The problem is that all the members of the ensemble are equally-likely (since we do not know the actual initial conditions), and thus they are indistinguishable from each other from a mathematical point of view. 

There is another viewpoint, however. While in dynamical, mathematical models all the ensemble members are equiprobable, real physical nonlinear systems have enough degrees of freedom to adjust itself to those trajectories compatible with the constraints of the system  \cite{Bag00} \cite{Breymann96}. We cannot observe such adjustment in Ulam's map or in other purely numerical dynamical systems, but we certainly can in physical systems \cite{Rebhan90}  \cite{Remler86} \cite{Gerard90} \cite{Kleidon03} \cite{Lorenz01} \cite{Dewar05}. 

This observation prompted \cite{Tapiador06} to propose a method for ranking members. Further refinement of that technique resulted in that it is possible to select a single member reducing as much as possible the error covariance while conserving a dynamically-consistent forecast, which is the method presented here. 

What any selection or blending procedure must ensure is that the result is both dynamically-consistent and the one surveying the $\rho (x)$ of the system. Ideally, it would be the less-biased choice. To find a less-biased member compatible with the constraints of the system, let us define a transient probability density function (pdf) of an ensemble member $j$ in phase space at $t$ as:
\begin{equation}
p_j^{t}(x)=\lim_{N  \rightarrow {\infty}} \frac{1}{N} \sum_{i=1}^{N} I_{x}^{t}[x] 
\end{equation}
where $I_{x}^{t}$ is the indicator function that gives 1 if the phase-space point of the trajectory $x$ is in the bin $(x-\delta,x+\delta )$ at time $t$, and 0 otherwise. For ergodic maps, the natural invariant density coincides with this pdf. In the general case, however, the time average does not have to be equal to the ensemble average and this pdf may depend on the ICs. 

We calculate the entropy of the pdf for each ensemble member at resolution $\delta$ as:
\begin{equation}
S_j^{t} \equiv  - \sum_{i=min}^{i=max} {p_{i,j}^{t}(x) \cdot log(p_{i,j}^{t}(x))}
\end{equation}
This entropy characterizes each member of the ensemble at time $t$. Crucially, the entropy in Eq. 6 is maximum for the flattest pdf compatible with the dynamics, and minimum for an hypothetical stationary system which orbit would remain in one bin. The difference between this estimate and the variance as pdf estimate is that the variance measures the concentration of the estimate only around the mean, whereas the entropy measures the spread of the pdf irrespective of the mode location. Also, a Legendre series expansion reveals that entropy is related to high-order moments of the pdf, unlike the variance.  

If we now considering the system generating the future climate pdf as a nonlinear physical system subject to a set of constraints, the more likely state of the system at $t$ is the one with the maximum entropy subject to those constraints \cite{Dewar03} \cite{Ozawa03}. In terms of the ensemble method, the member having the maximum entropy will display the most wandering trajectory in phase space since by construction it represents the broadest distribution compatible with the constraints. For those pdfs similar to the natural invariant density of Ulam's map (Figure 1) the maximum entropy member would be the one more evenly surveying the $[-1,1]$ interval. In other words, the maximum entropy member in this case would be the member compatible with the dynamics whose trajectory more evenly preserves the bimodal pdf of the map. 

The conclusion is that the maximum entropy ensemble member should be the member to compare with experiments in climate models, because, as in Statistic Mechanics, the ensemble average might be the average of many peaks and itself correspond to an impossible value \cite{Jaynes57}. It is important to note that since we do not have evidence to decide for any member in particular all we can do is aiming for the least worse member, which is the one maximizing the entropy. In other words, the procedure is not about finding the best member but about using a method that ensures that we are committing the least possible error when selecting one. 

For unimodal distributions, the maximum entropy member corresponds with the member with the highest variance: if the pdf is normally distributed, the relationship between the entropy and the variance is $S= -\sum_{i} p(i) \cdot log (p(i)) = log(\sigma \sqrt{2 \pi e})$ with $\sigma$ the standard deviation. For the lognormal pdf, $ S= \frac{1}{2} + \frac{1}{2} log (2 \pi \sigma^{2})$. (This relationships hold for the whole phase space, so they are not to be confused with the daily results in Table 1). For non-analytical pdfs, the selection procedure requires the actual calculation of the entropy of the ensemble members as the relationship between entropy estimate and the variance cannot be analytically described. 

As the actual trajectory of the system in phase space will be one in a maximum entropy state, the member with the highest entropy will be the closest to the true state. By construction, the error covariance of the maximum entropy member compared with the true trajectory is automatically the minimum error covariance of the whole ensemble.

To test this hypothesis we have used more cases included the DEMETER project. Even having this validation data, the issue of how to establish which member of the 180-days forecast meteorological ensemble is the best is not trivial \cite{Elmore05} \cite{Muller05}: the objective best member may change for one day to the  next, without a clear overall winner \cite{Roulston03}. To minimize this problem, the quantity $\left\langle {r}^2 \right\rangle _T = \frac{1}{T}\sum_{t=1}^T r_t^{2}$, ($r^2$ being the correlation coefficient) can be used as goodness-of-fit metric. We have considered that the best forecast for day $T$ is not the one in which the forecast compares better with the experimental data, but the one with the best averaged goodness-of-fit from the first day of the forecast. Thus, the best member is defined as the member that is closest to the validation in the full state space of the model. Therefore we have computed $max \left\{ \left\langle {S_j} \right\rangle _T \right\}$ (Eq. 6), and compare with $max \left\{ \left\langle {r_j}^2 \right\rangle _T \right\}$ for each member $j$. 

What we found is that the entropy estimate relates with the best member. Figure 5 shows the evolution of the entropy choice and the best and worse objective members for a set of five, 180 days, 9 ensemble-members runs. After the spin-up of the model, the maximum entropy member is consistently closest to the objective best member than to the worse member. These results are stronger than those reported in \cite{Tapiador06} making unnecessary a weighted average of the members.

\section{Conclusions}
Ensemble procedures provide a way to palliate SIC in nonlinear dynamical systems in general, and climate models in particular. Averaging of climate ensemble forecast, however, generates an unphysical blended product, reducing the variance of the estimate and forecasting a spurious pdf for the future climate. The immediate consequence is that member-averaged climate forecasts severely underestimate climatological extremes such as high precipitation rates. 

The alternative to averaging is to select a single member of the ensemble that both preserves the physical character of the forecast and maintains the pdf structure. In absence of any evidence to select a priori equiprobable member, we should lean toward the maximum entropy member. This member  differs from other members in representing the most likely trajectory of the system which is compatible with the constraints. Therefore, it is the least worse choice with the available information. In a problem subject to SIC such as Earth's climate modeling, this least assumptive feature is an important characteristic that climate forecasts should seek to avoid underestimating climatological extremes.


\clearpage

\begin{figure}
\begin{center}
\includegraphics[width=14cm]{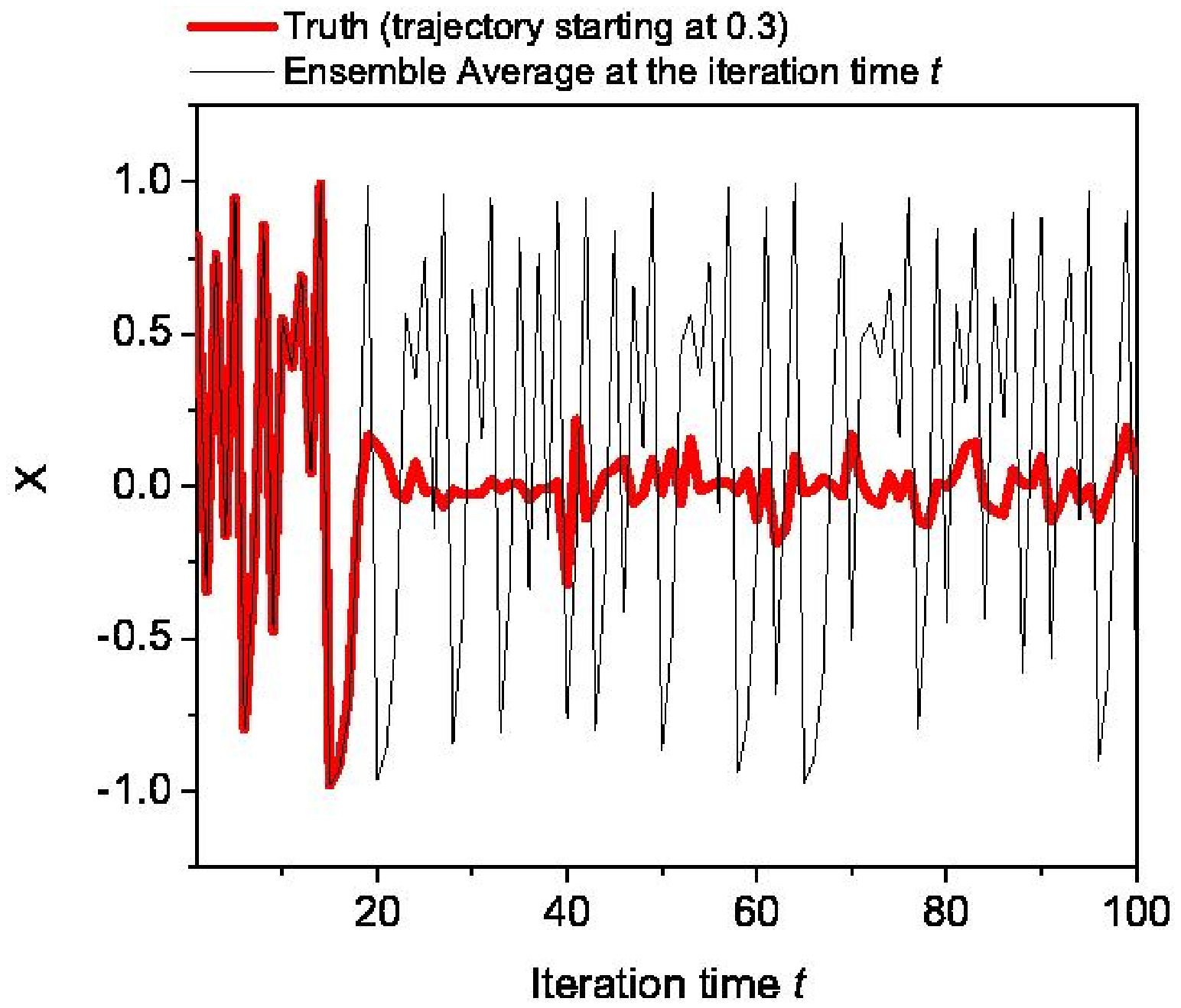}
\caption{\label{fig} Ulam's map trajectories for a typical member of the ensemble (orbit starting at $x_0=0.3$, thin line) and the ensemble average (bold line). The ensemble is composed by 50 members starting within $[x_0-10^{-8}, x_0+10^{-8}]$. After 20 iterations, the trajectories start to differ.}
\end{center}
\end{figure} 
\clearpage


\begin{figure}
\begin{center}
\includegraphics[width=12cm]{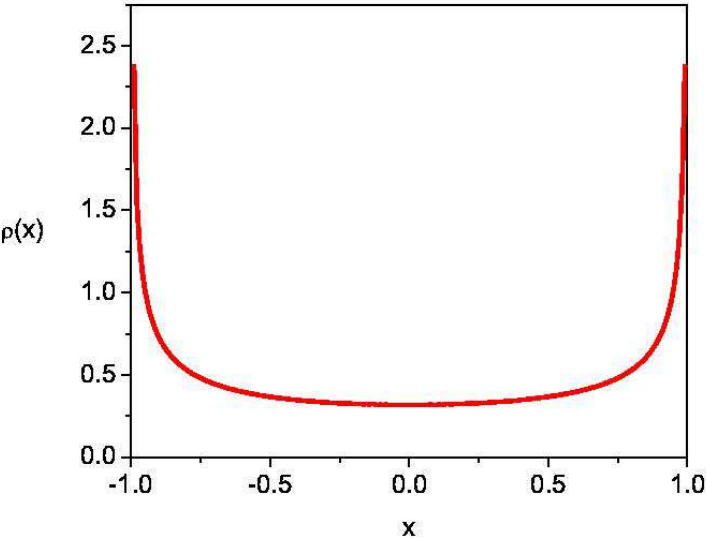}
\caption{\label{fig} The natural invariant density of Ulam's map} 
\end{center}
\end{figure} 
\newpage
\clearpage
\clearpage


\begin{figure}
\begin{center}
\includegraphics[width=12cm,angle=90]{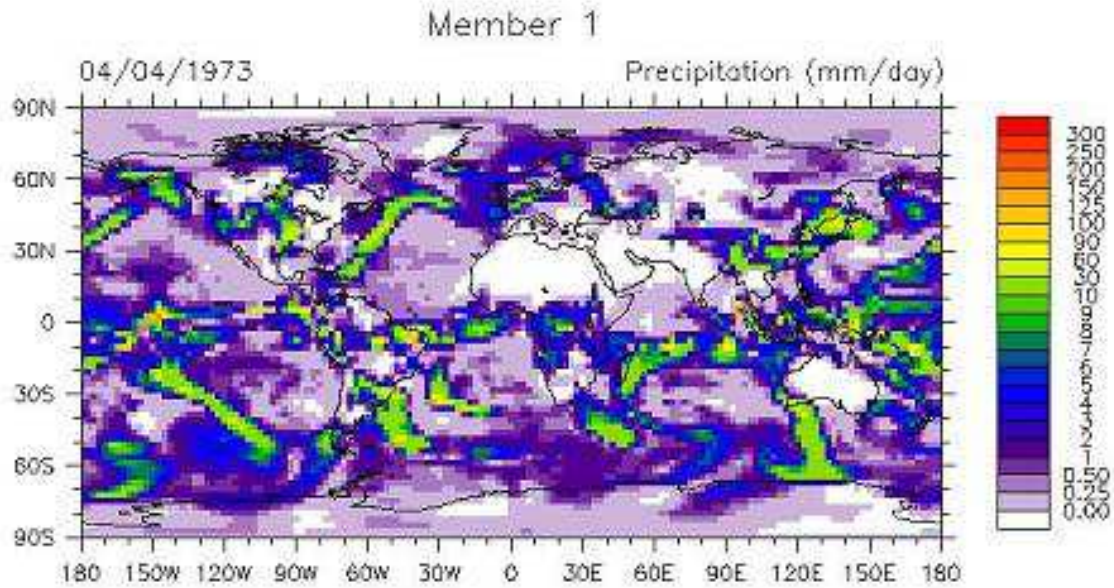}
\caption{\label{fig} (1/11) Precipitation forecast for April 4th, 1973, comparing validation data (ERA-40 reanalysis) with the 9 members of the DEMETER ensemble and with the ensemble average. The ensemble average provides a slightly better correlation than the objective best member (which is member 6, cfr. table I) at the cost of providing a forecast which is incompatible with the dynamics.}
\end{center}
\end{figure} 
\clearpage

\begin{figure}
\addtocounter{figure}{-1}
\begin{center}
\includegraphics[width=12cm,angle=90]{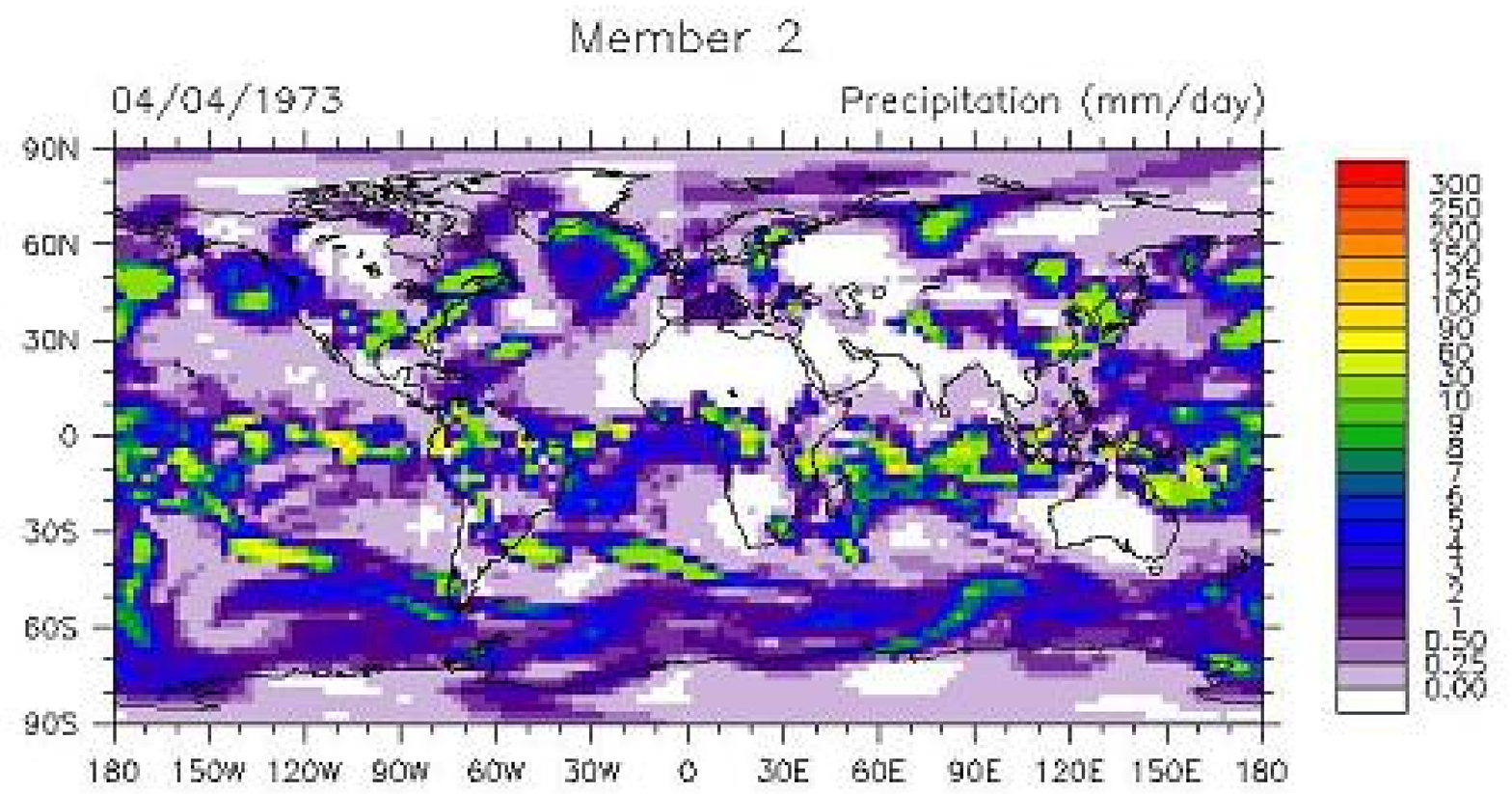}
\caption{\label{fig} cont (2/11) }
\end{center}
\end{figure}
\clearpage

\begin{figure}
\addtocounter{figure}{-1}
\begin{center}
\includegraphics[width=12cm,angle=90]{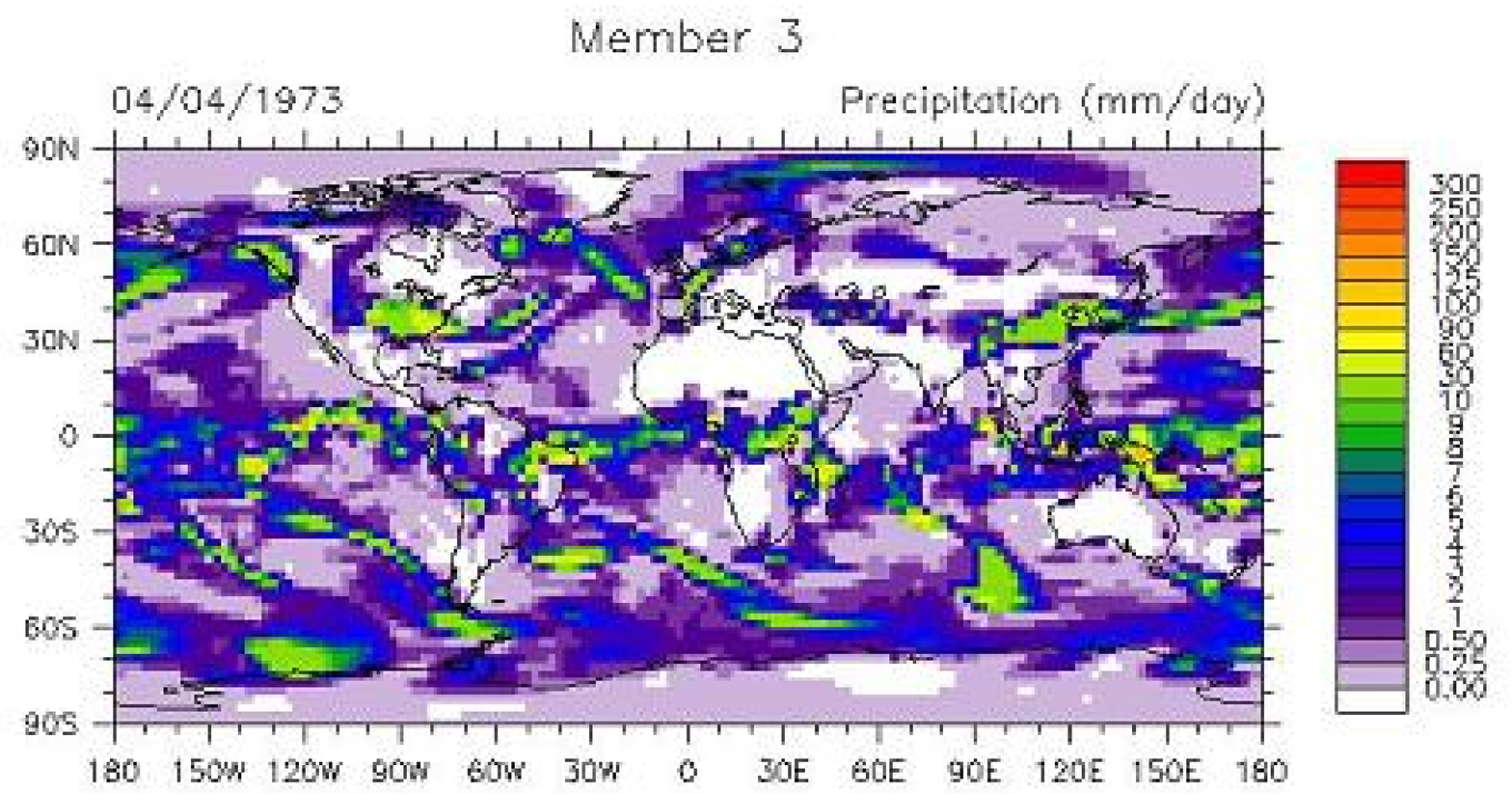}
\caption{\label{fig} cont (3/11) }
\end{center}
\end{figure}
\clearpage

\begin{figure}
\addtocounter{figure}{-1}
\begin{center}
\includegraphics[width=12cm,angle=90]{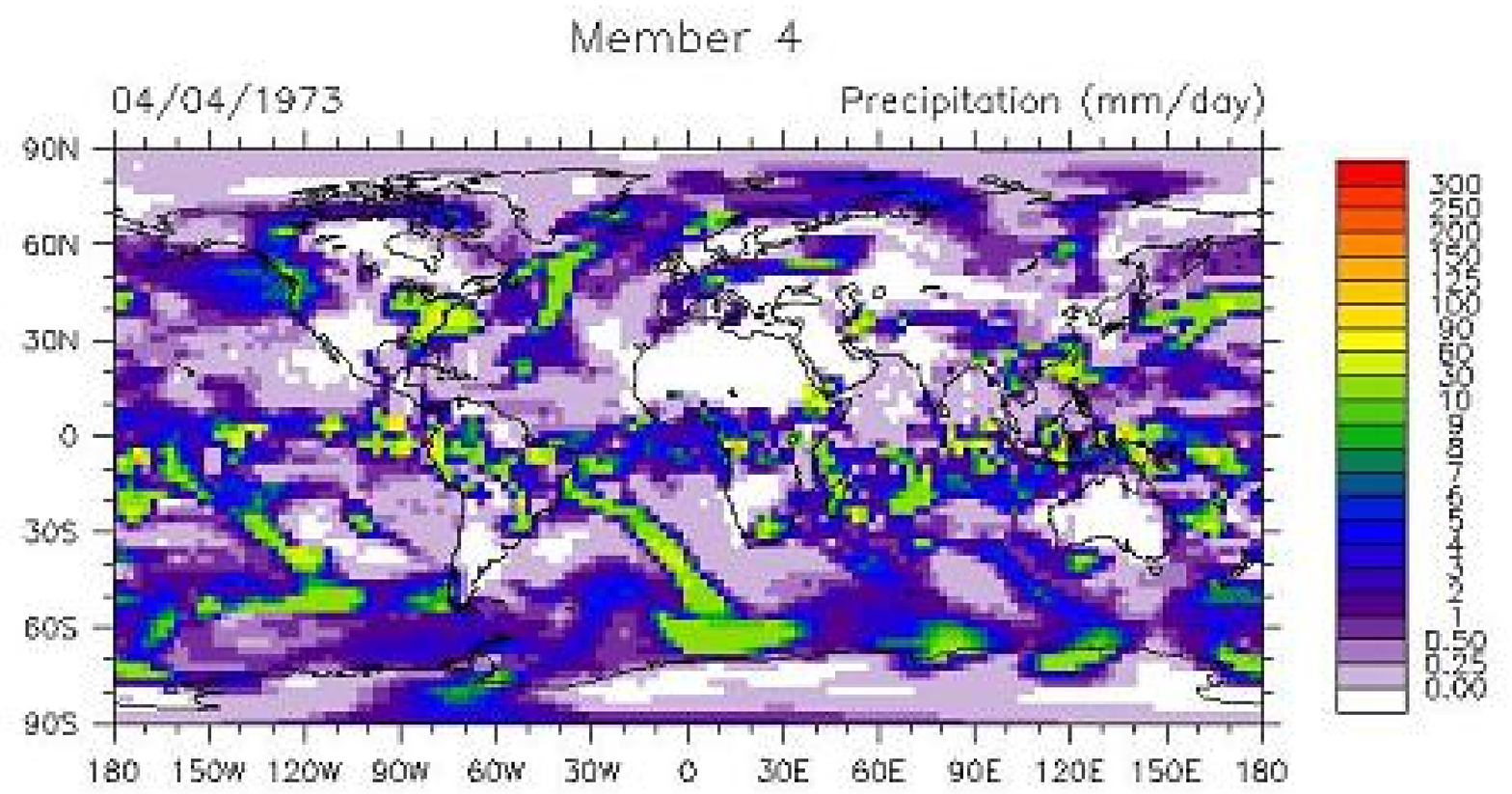}
\caption{\label{fig} cont (4/11) }
\end{center}
\end{figure}
\clearpage

\begin{figure}
\addtocounter{figure}{-1}
\begin{center}
\includegraphics[width=12cm,angle=90]{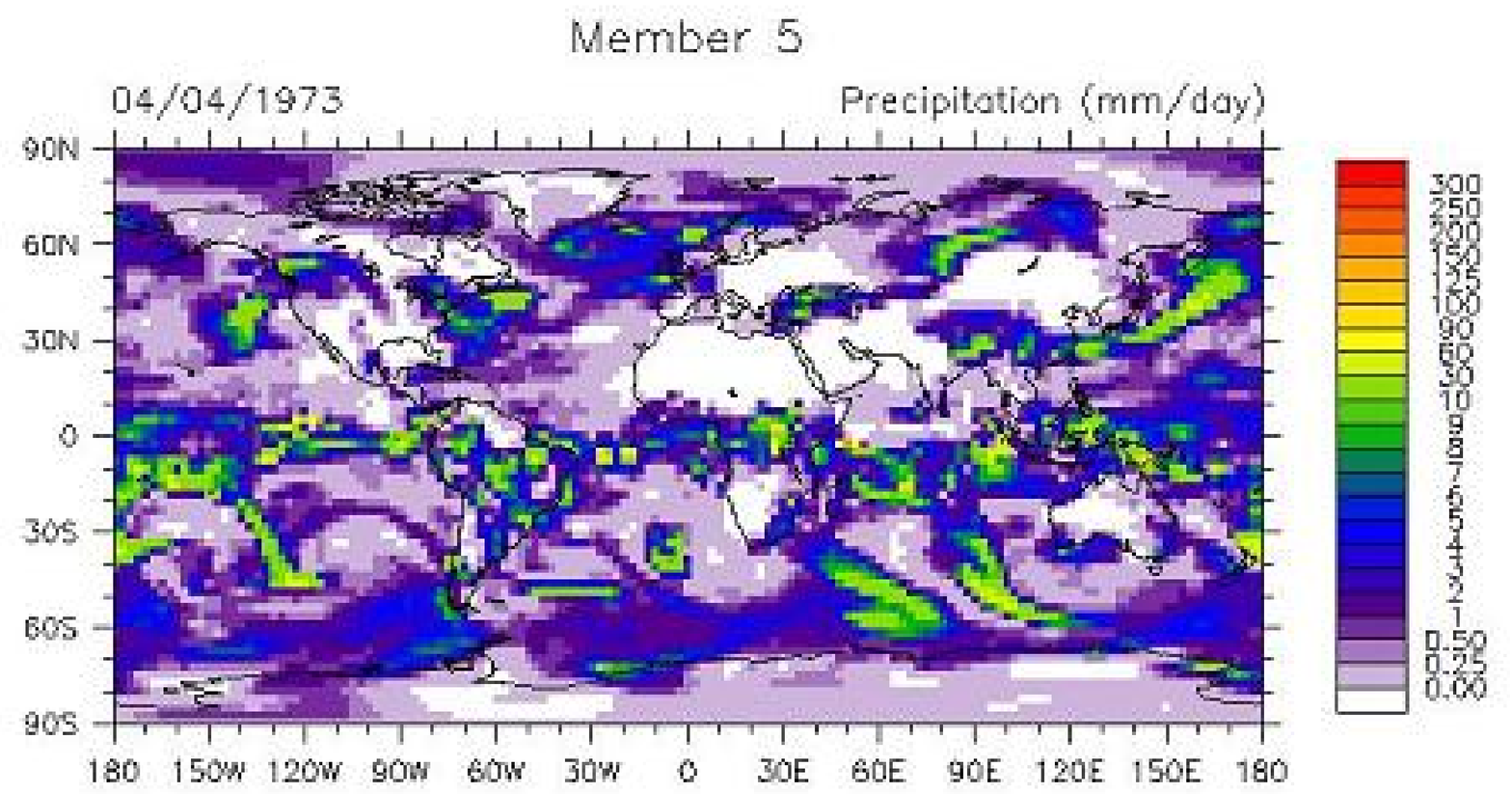}
\caption{\label{fig} cont (5/11) }
\end{center}
\end{figure}
\clearpage

\begin{figure}
\addtocounter{figure}{-1}
\begin{center}
\includegraphics[width=12cm,angle=90]{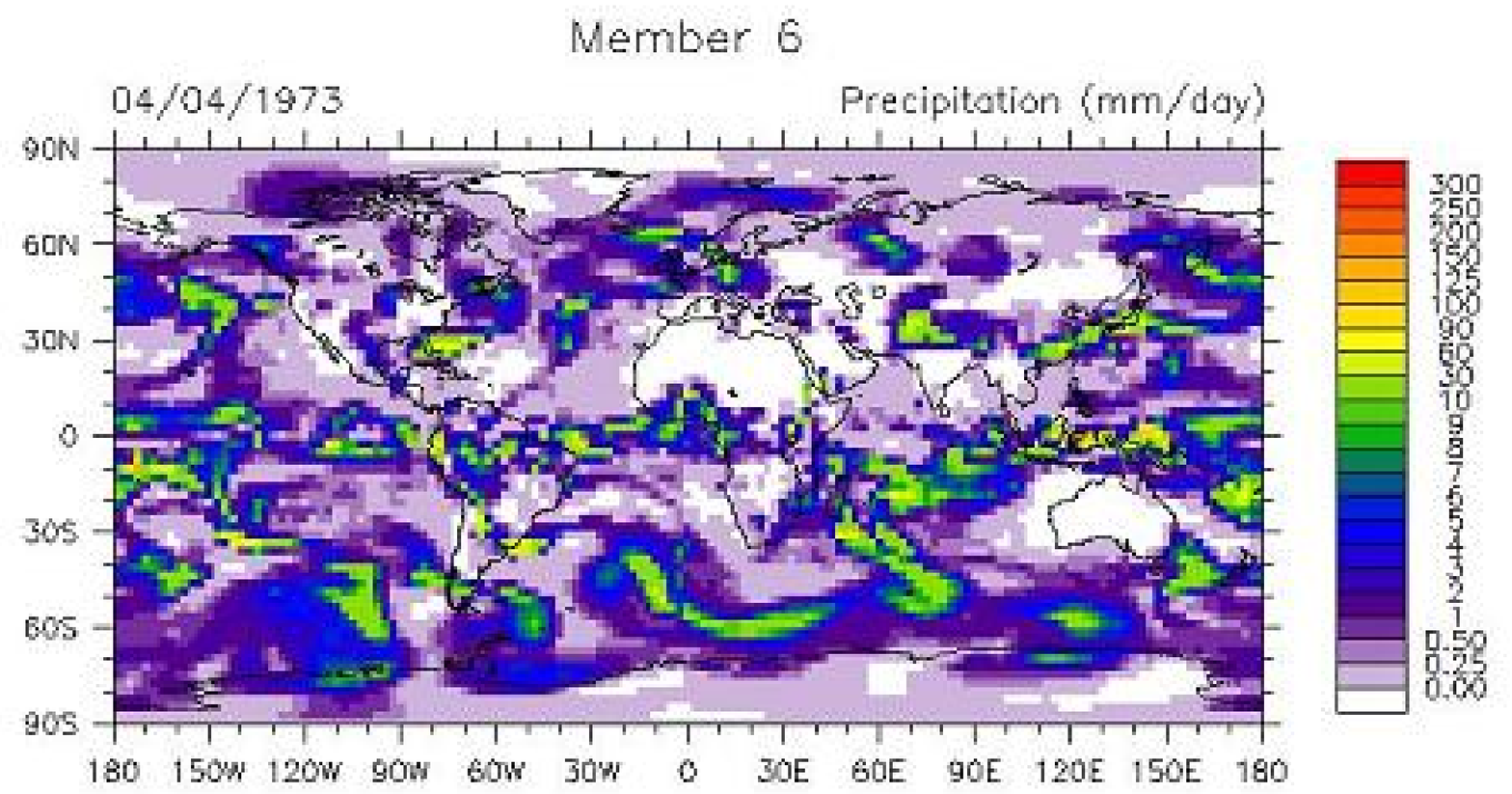}
\caption{\label{fig} cont (6/11) }
\end{center}
\end{figure}
\clearpage

\begin{figure}
\addtocounter{figure}{-1}
\begin{center}
\includegraphics[width=12cm,angle=90]{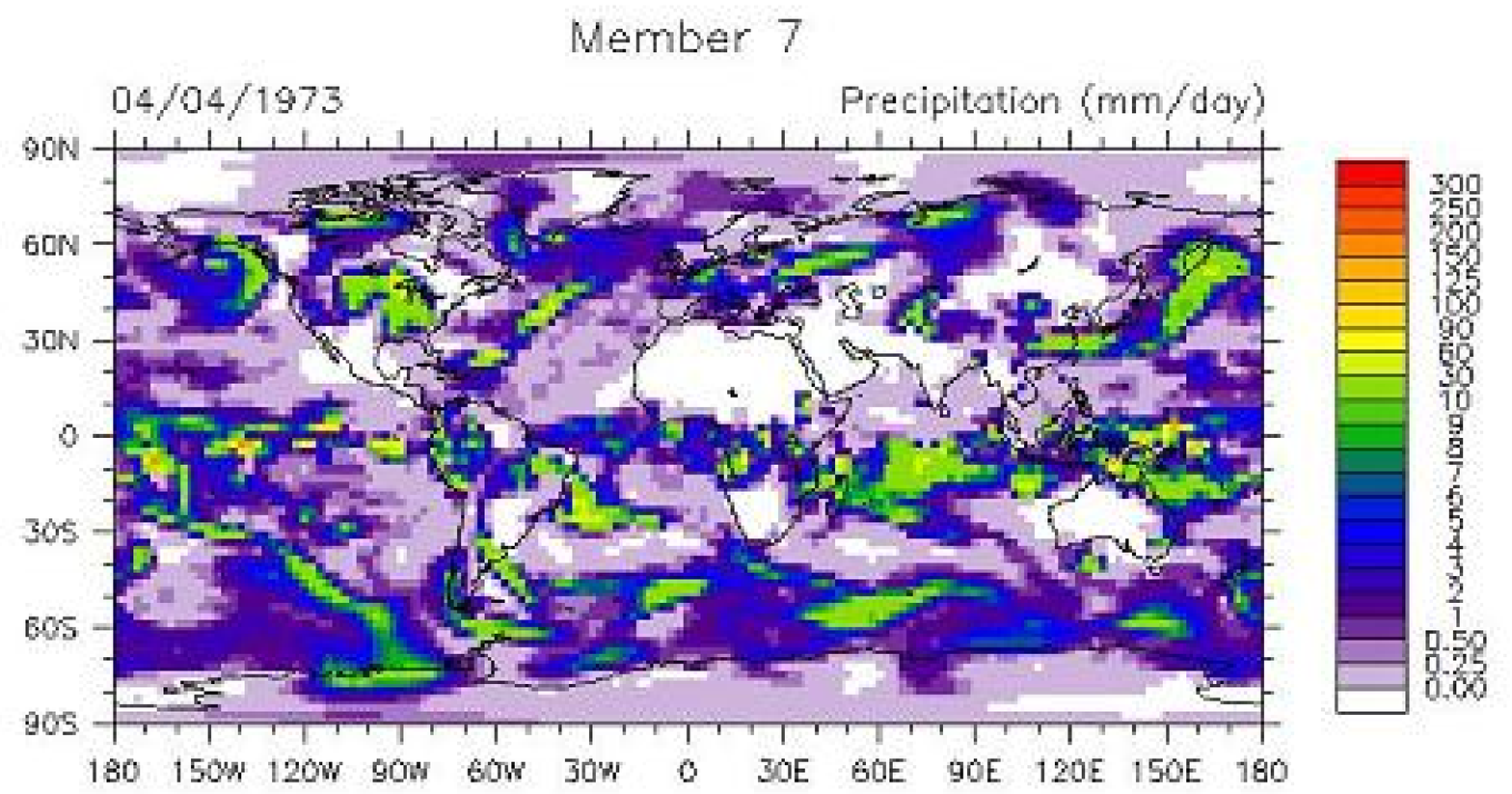}
\caption{\label{fig} cont (7/11) }
\end{center}
\end{figure}
\clearpage

\begin{figure}
\addtocounter{figure}{-1}
\begin{center}
\includegraphics[width=12cm,angle=90]{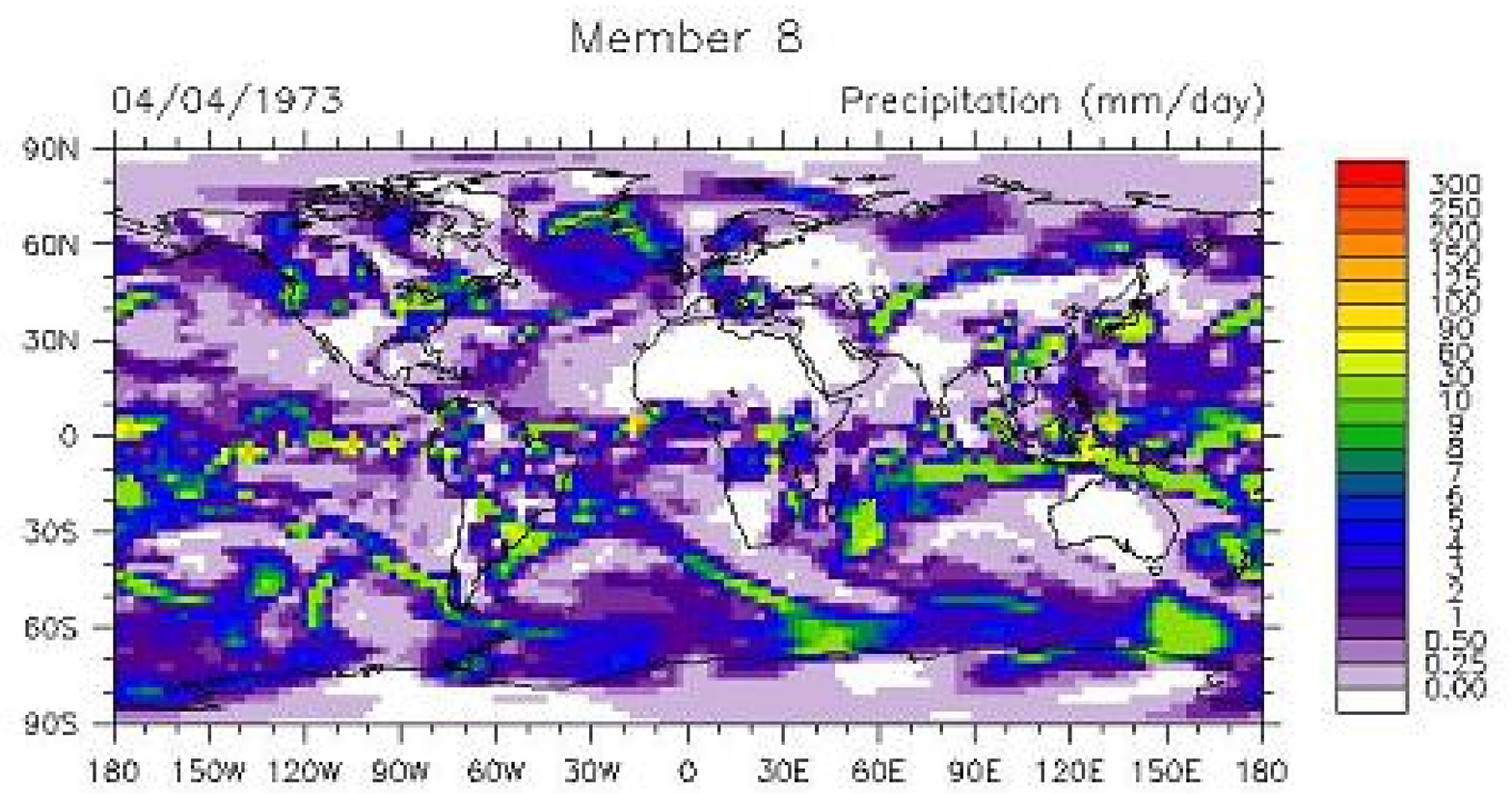}
\caption{\label{fig} cont (8/11) }
\end{center}
\end{figure}
\clearpage

\begin{figure}
\addtocounter{figure}{-1}
\begin{center}
\includegraphics[width=12cm,angle=90]{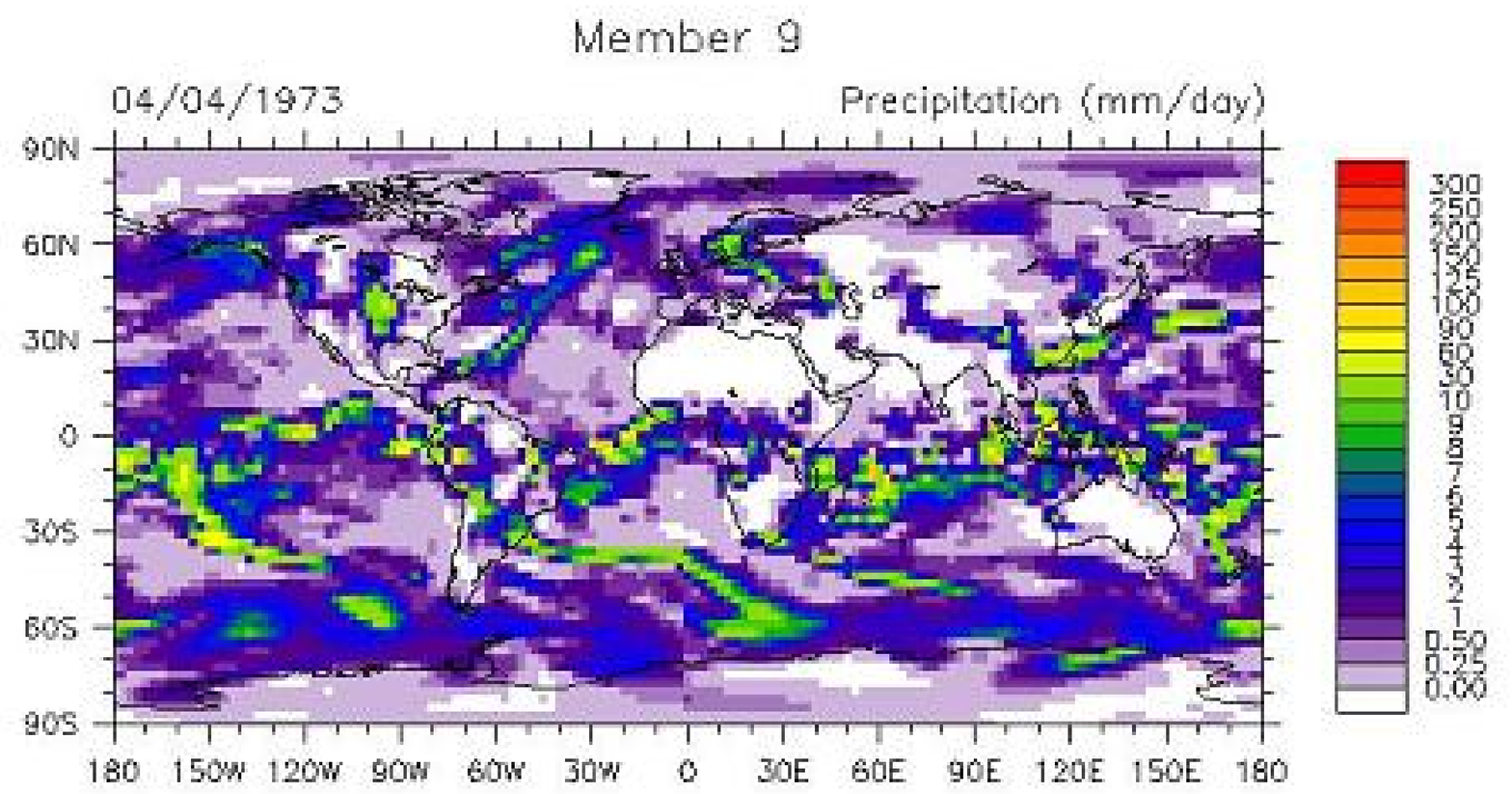}
\caption{\label{fig} cont (9/11) }
\end{center}
\end{figure}
\clearpage

\begin{figure}
\addtocounter{figure}{-1}
\begin{center}
\includegraphics[width=12cm,angle=90]{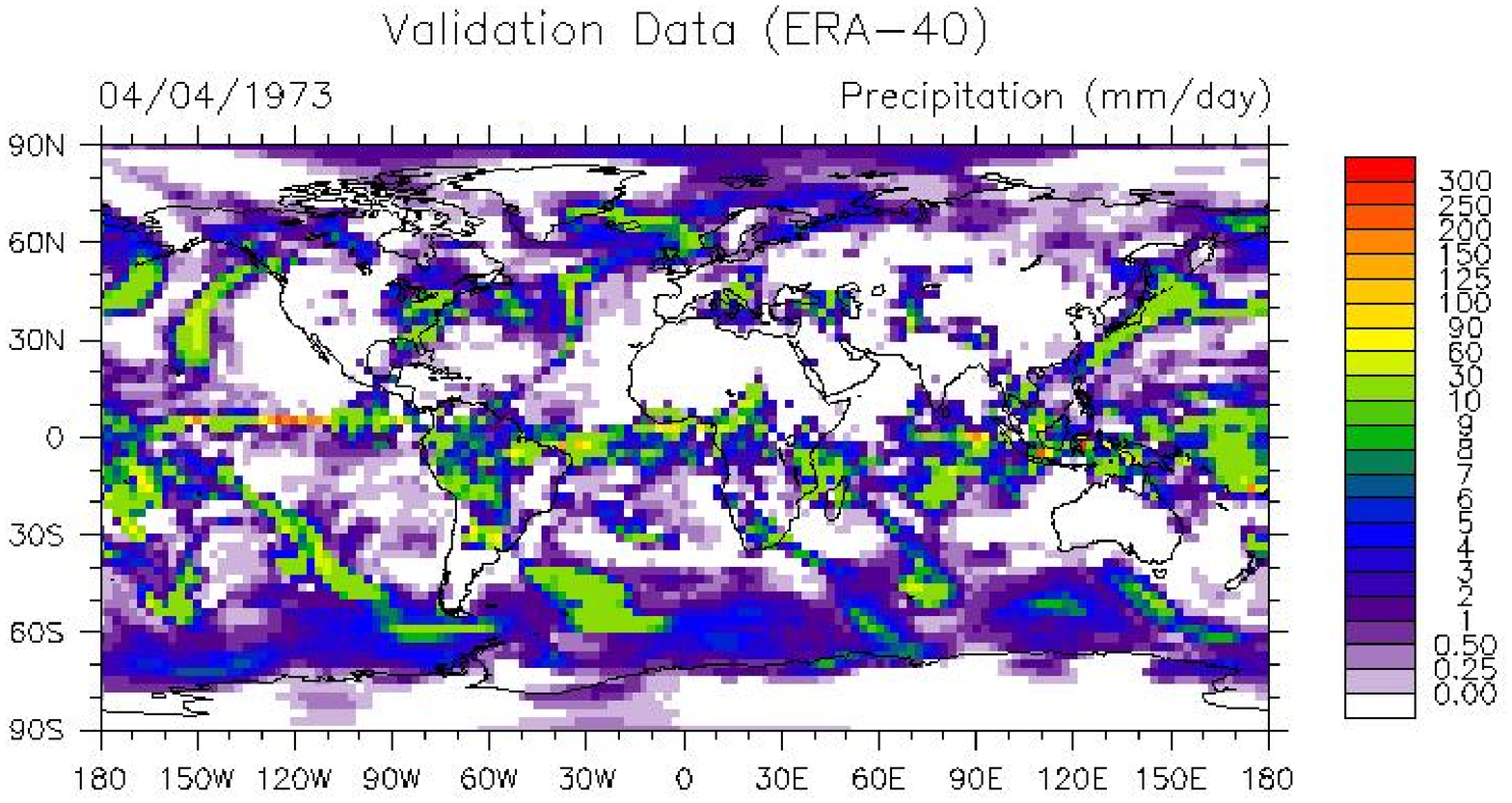}
\caption{\label{fig} cont (10/11) }
\end{center}
\end{figure}
\clearpage

\begin{figure}
\addtocounter{figure}{-1}
\begin{center}
\includegraphics[width=12cm,angle=90]{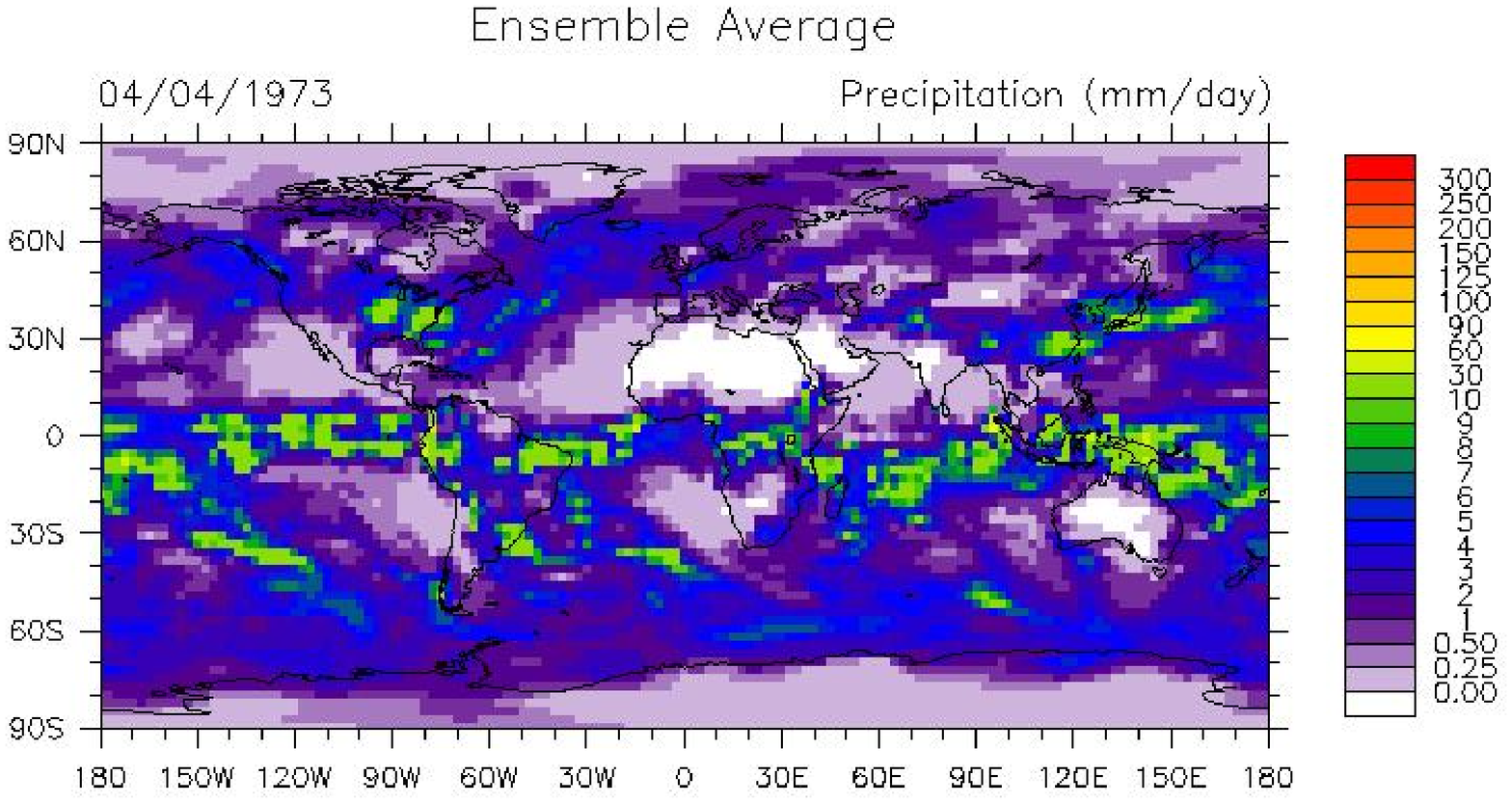}
\caption{\label{fig} cont (11/11) }
\end{center}
\end{figure}
\clearpage

\begin{figure}
\begin{center}
\includegraphics[width=12cm]{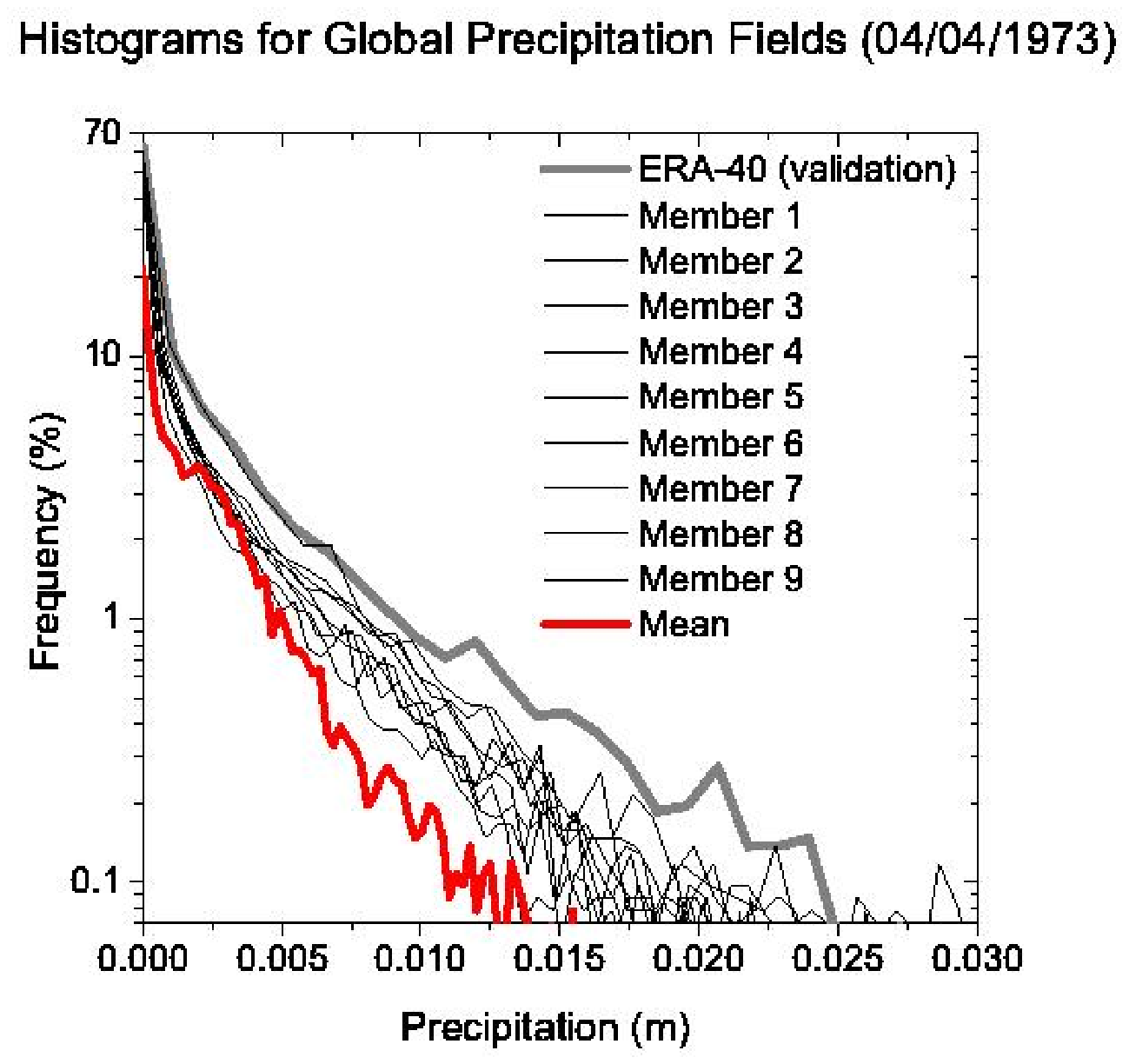}
\caption{\label{fig} Histograms of the precipitation estimates in figure 3. Y-axis in logarithmic scale, X-axis truncated at 0.03m of precipitation}
\end{center}
\end{figure}
\clearpage

\begin{figure}
\begin{center}
\includegraphics[width=14cm,angle=0]{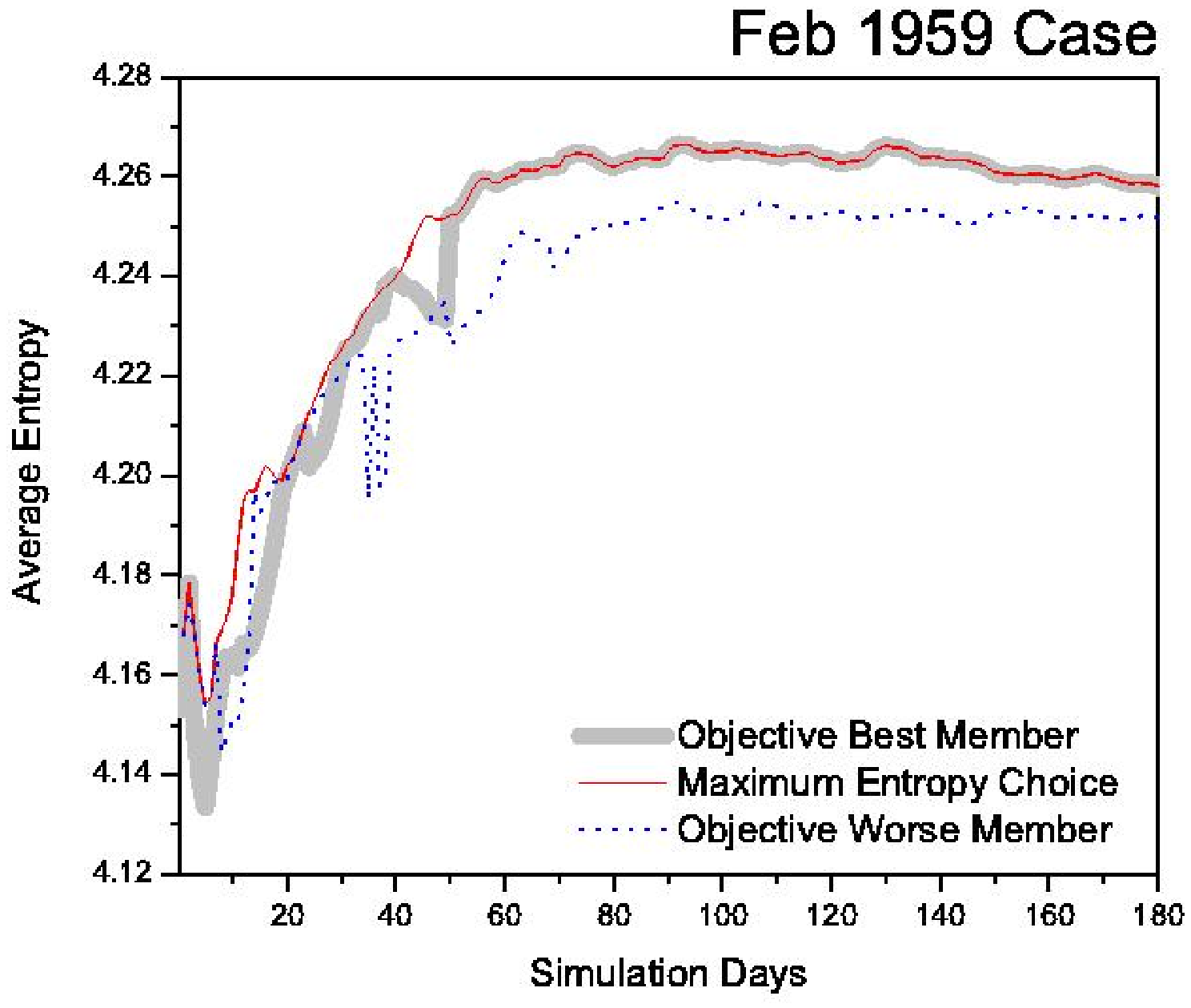}
\caption{(1/5)Performance of the entropy selection method for four 180-days forecasts (DEMETER data). The maximum entropy choice is consistently close (resp. further) to the best (resp. worse) objective members}
\end{center}
\end{figure}
\clearpage

\begin{figure}
\begin{center}
\addtocounter{figure}{-1}
\includegraphics[width=14cm,angle=0]{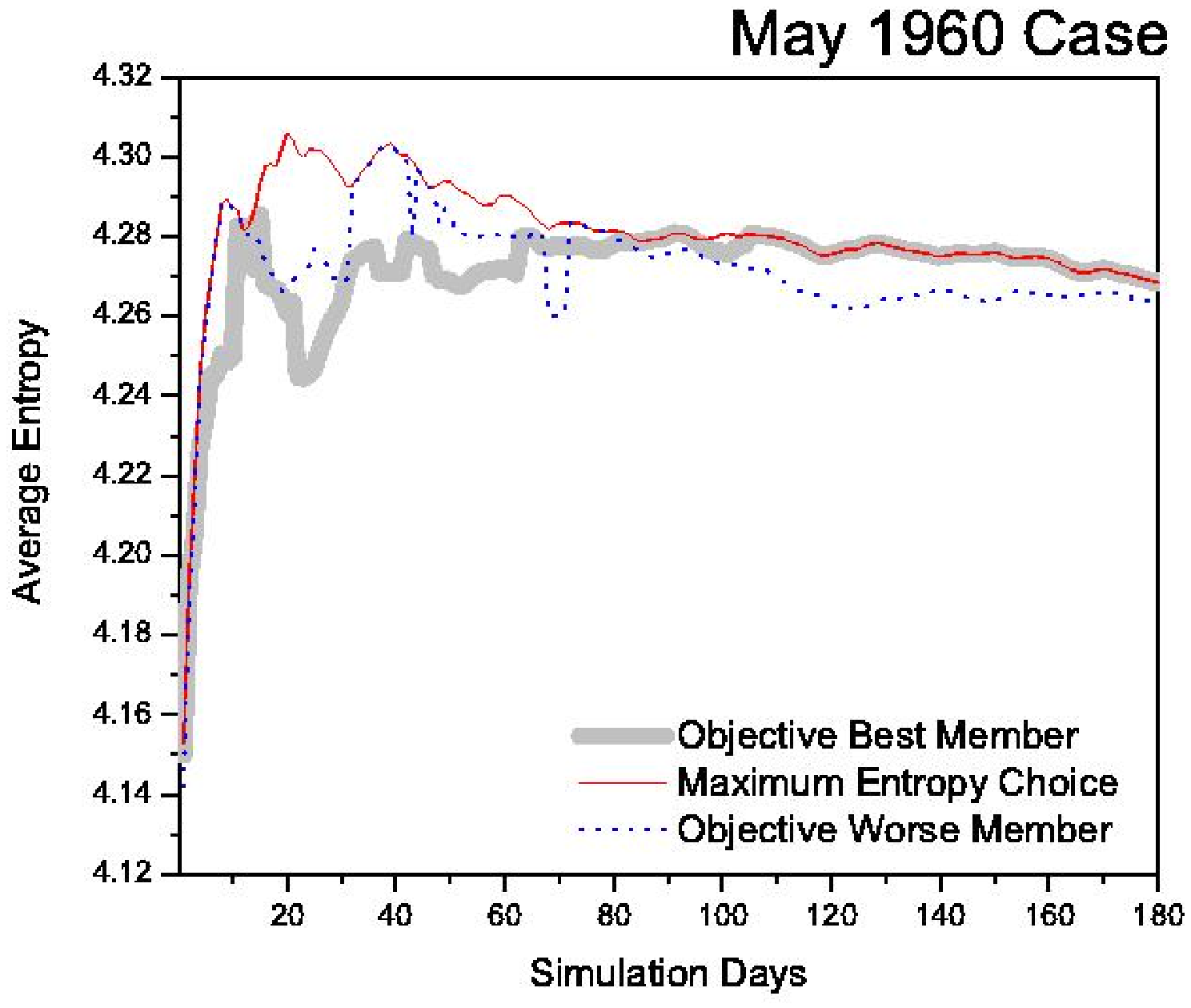}
\caption{cont (2/5)}
\end{center}
\end{figure}
\clearpage

\begin{figure}
\begin{center}
\addtocounter{figure}{-1}
\includegraphics[width=14cm,angle=0]{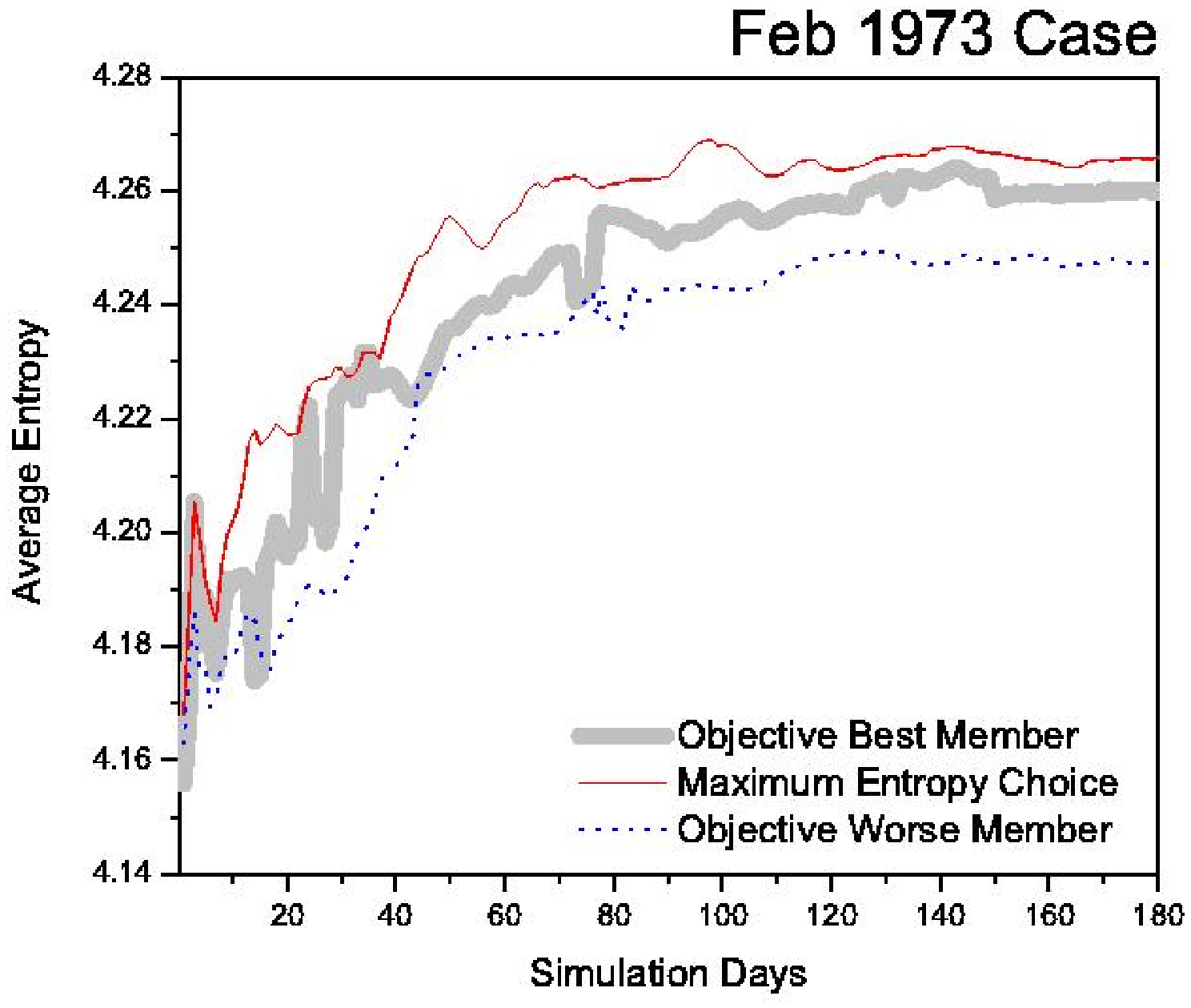}
\caption{cont (3/5)}
\end{center}
\end{figure}
\clearpage

\begin{figure}
\begin{center}
\addtocounter{figure}{-1}
\includegraphics[width=14cm,angle=0]{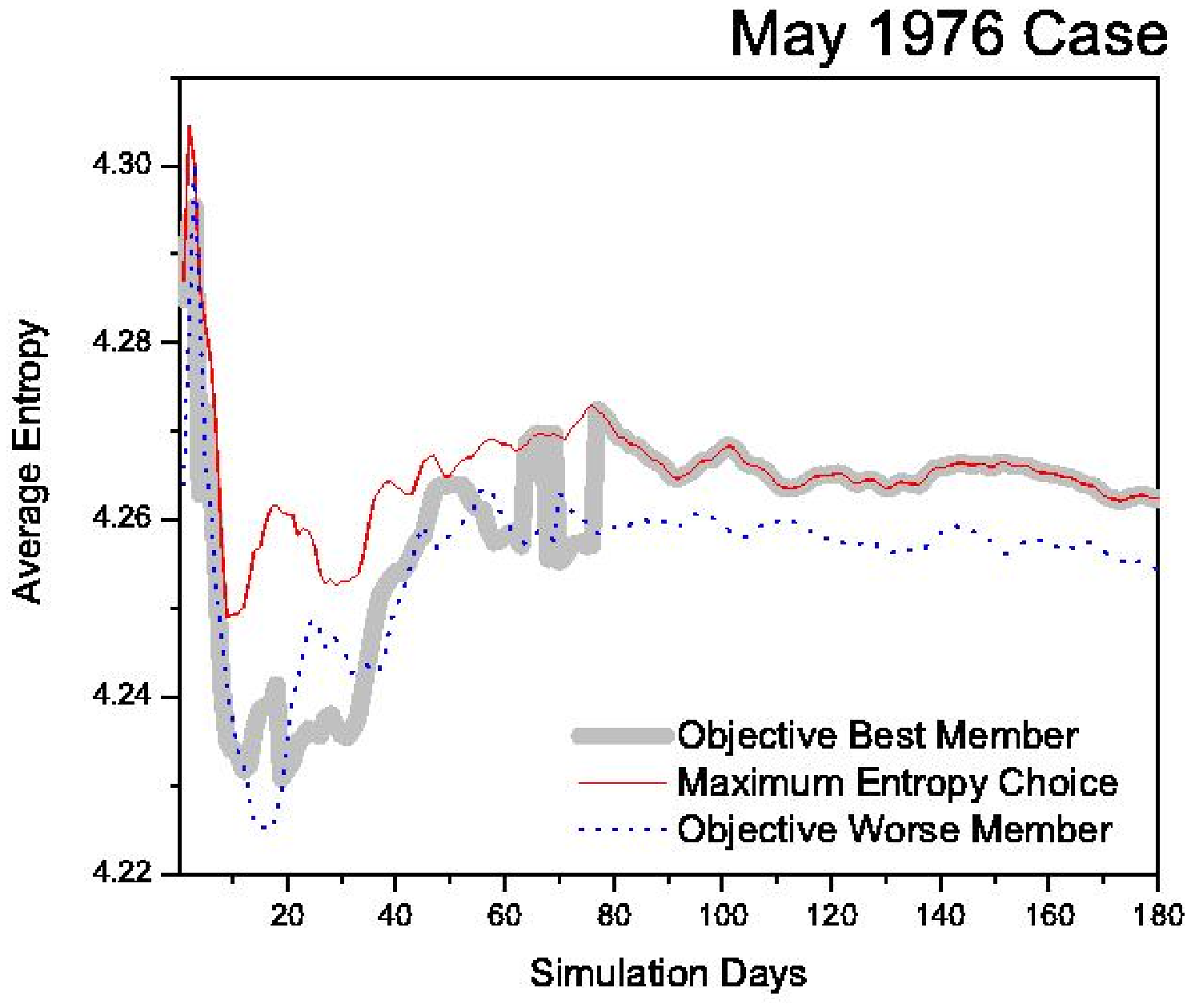}
\caption{cont (4/5)}
\end{center}
\end{figure}
\clearpage

\begin{figure}
\begin{center}
\addtocounter{figure}{-1}
\includegraphics[width=14cm,angle=0]{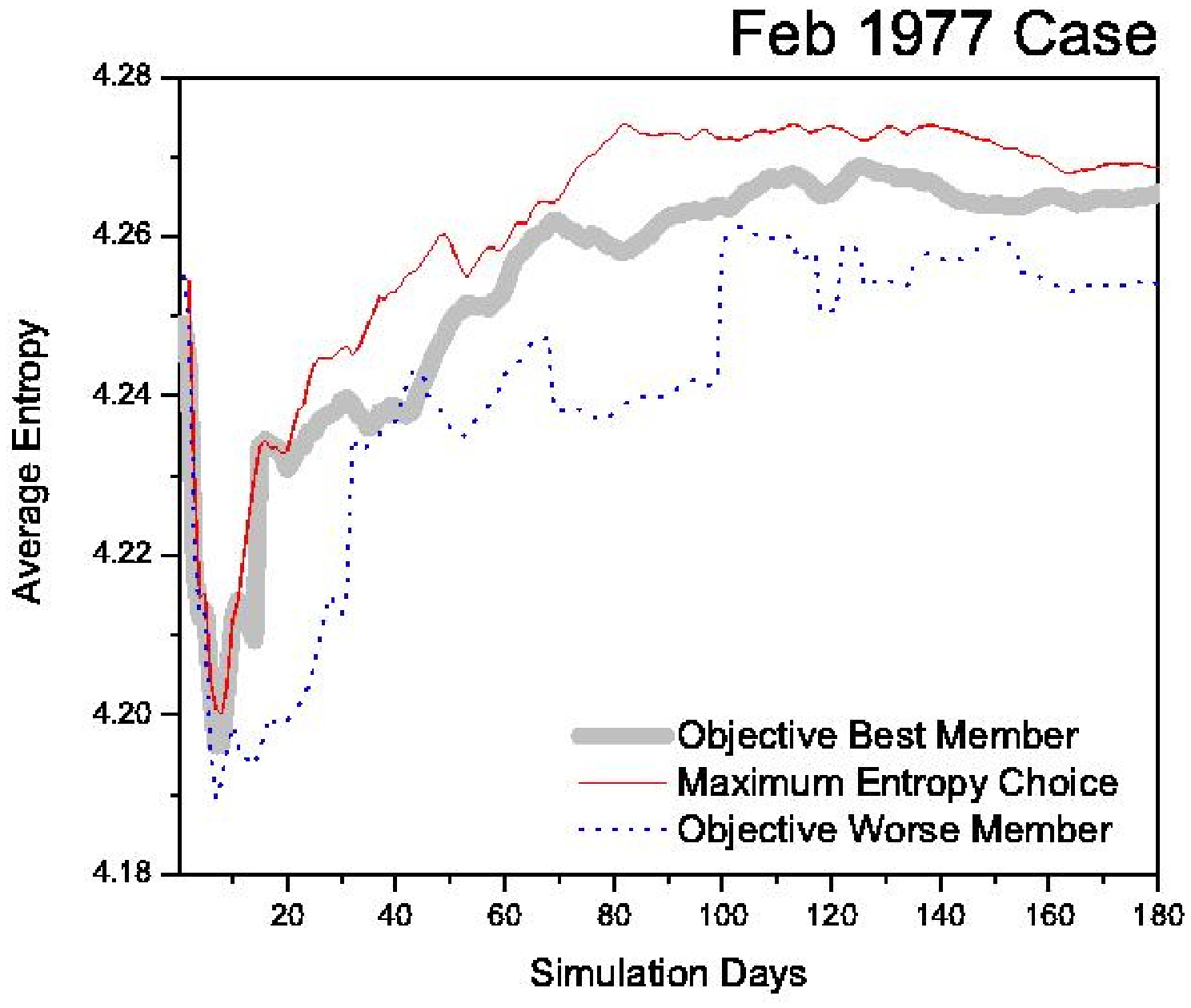}
\caption{cont (5/5)}
\end{center}
\end{figure} 
\clearpage
\begin{table}
\begin{center}
\caption{\label{tab:table1}Statistics for April 4th, 1973, featuring the correlation ($r^2$), the maximum (mm/day), mean (mm/day) and standard deviation (mm/day) values of the precipitation forecast fields. ERA-40 reanalysis data is deemed as the truth (validation data).}
\begin{tabular}{lcrll}
 & $r^2_{Precip}$ & $max_{Precip}$ & $\mu_{Precip}$ & Spatial $\sigma_{Precip}$ \\
\hline 
Member 1 & 0.1053 &  187.069 & 2.516 & 6.964      \\
Member 2 & 0.0949 &  116.356 & 2.354 & 6.516      \\
Member 3 & 0.1111 &  147.500 & 2.380 & 6.742      \\
Member 4 & 0.1053 &  145.808 & 2.592 & 6.776      \\
Member 5 & 0.2075 &  153.411 & 2.194 & 5.340      \\
Member 6 & 0.1399 &  193.796 & 2.258 & 6.280      \\
Member 7 & 0.0894 &  165.628 & 2.454 & 6.677      \\
Member 8 & 0.0911 &  246.747 & 2.324 & 6.530     \\
Member 9 & 0.1087 &  139.182 & 2.248 & 6.371     \\
Truth &   (1.0000) & 277.449 & 2.744 & 9.174  \\
Ensemble avg &0.2247 &  62.509 &  2.369 & 3.307 \\   
\end{tabular}
\end{center}
\end{table}
\clearpage


\end{document}